\documentclass[preprintnumbers,superscriptaddress,showkeys,showpacs,byrevtex]{revtex4}   
\usepackage{amsmath,amsfonts,amssymb,amscd,amsxtra,amsthm}
\usepackage{graphicx}
\usepackage{bm}
\usepackage{epstopdf}
\usepackage{multirow}
\begin{document}  
\preprint{KIAS-P11048}
\preprint{CYCU-HEP-11-15}
\title{Unpolarized fragmentation function for the pion and kaon\\
via the nonlocal chiral-quark model}      
\author{Seung-il Nam}
\email[E-mail: ]{sinam@kias.re.kr}
\affiliation{School of Physics, Korea Institute for Advanced Study (KIAS), Seoul 130-722, Republic of Korea}
\author{Chung-Wen Kao}
\email[E-mail: ]{cwkao@cycu.edu.tw}
\affiliation{Department of Physics, Chung-Yuan Christian University (CYCU), Chung-Li 32023, Taiwan}
\author{Byung-Geel Yu}
\email[E-mail: ]{bgyu@kau.ac.kr}
\affiliation{Research Institute for Basic Sciences, Korea Aerospace University (KAU), Goyang 412-791, Republic of Korea}
\date{\today}
\begin{abstract}
In this talk we present our recent studies for the unpolarized fragmentation functions for the pion and kaon, employing the nonlocal chiral quark model, which manifests the nonlocal interaction between the quarks and pseudoscalar mesons, in the light-cone frame. It turns out that the nonlocal interaction produces considerable differences in comparison to typical local-interaction models. 
\end{abstract} 
\maketitle
\section{Introduction}
Recently, there has been an increasing interest for the nonperturbative quantities of quantum chromodynamics (QCD), observed in hard scattering processes, together with the experimental and theoretical progresses. Among those physical quantities, fragmentation and parton-distribution functions reveal important information on the internal structures of the hadrons.  Especially, the fragmentation function plays an crucial role in describing semi-inclusive hadron productions in electron annihilation and deeply inelastic scattering (DIS) processes. 

In Ref.~\cite{Hirai:2007cx}, various fragmentation functions are intensively investigated using the $Q^2$ evolution equation, such as the Dokshitzer-Gribov-Lipatov-Altarelli-Parisi (DGLAP) equation, resulting in empirical parametrizations for those fragmentation functions. More practical approaches have been done for the fragmentation functions, taking into account the effective chiral models for QCD, i.e. liner (pseudoscalar, PS) and nonlinear (pseudovector, PV) $\sigma$ models (or quark-pseudoscalar (PS) meson coupling model)~\cite{Bacchetta:2002tk,Amrath:2005gv}, and Nambu-Jona-Lasinio (NJL) model~\cite{Ito:2009zc}. It is worth noting that, in Refs.~\cite{Bacchetta:2002tk,Amrath:2005gv}, it was not uniquely determined which coupling schemes, i.e.  PS and PV ones, are more reliable to describe the fragmentation function. Comparing with the NJL model calculations~\cite{Ito:2009zc}, the PS scheme results looks similar to that from the NJL model. In Ref.~\cite{Ito:2009zc}, the authors performed the DGLAP evolution using the numerical code QCDNUM for the DGALP evolution~\cite{Botje:2010ay, DGLAP}, in addition to the contributions from the resonances and jets. 

In the present work, we employ the nonlocal chiral quark model (NLChQM) defined in Minkowski space. A unique feature of this model, which is not observed in other models, is that quarks and PS mesons are interacting each other nonlocally: This interesting nonlocal interaction can be regarded as the consequence of the quark and nontrivial QCD vacuum interactions in terms of the the semiclassical nontrivial solution for the Yang-Mills equation, i.e. instanton~\cite{Diakonov:1985eg,Shuryak:1981ff,Diakonov:1983hh,Diakonov:2002fq,Schafer:1996wv}. From the numerical calculations, we observe that the present model provides quite distinctive results in comparison to those from other models. Especially, the bump shown in the fragmentation function curve appears at $z\approx0.5$, whereas the PS-scheme and NJL results produce it at larger $z$ values $\sim0.7$. This tendency observed in the present framework, saying, the bump appears in the lower $z$ regions, may give more comparable results with the empirical data at higher $Q^2$ after the DGLAP evolution, in comparison to other models.  

The present report is structured as follows: In Sections II and III, we briefly introduce the unpolarized fragmentation function and the NLChQM which is based on the instanton vacuum configuration in principle. The numerical results are given in Section IV with related discussions. Final Section is devoted to summary, conclusion, and future plans.  

\section{Unpolarized fragmentation function}
The unpolarized fragmentation function $(d^\phi_{q,\bar{q}})$ for $q^*_f\to\phi\,q_{f'}$, in which $q_f$ and $\phi$ indicate the quark with the flavor $f$ and PS meson respectively, is defined in the light-cone coordinate, assuming the light-cone gauge, as follows~\cite{Bacchetta:2002tk,Amrath:2005gv,Mulders:1995dh}:
\begin{equation}
\label{eq:FRAG}
d^\phi_{q,\bar{q}}(z,k^2_T,\mu)=\frac{1}{4z}\int dk^+\mathrm{Tr}
\left[\Delta(k,p,\mu)\gamma^- \right]|_{zk^-=p^-},
\end{equation}
where $p$ and $z$ stands for the on-shell four momentum  of the meson and its momentum faction possessed by a quark fragmented, whereas $k_{\pm}$ for the time- and space-momentum components of the quark on the light cone, respectively. $k_T$ and $\mu$ denotes the transverse momentum of the quark and renormalization scale at which the fragmentation process evolved. For the later convenience, we assign the fragmentation function  in Eq.~(\ref{eq:FRAG}) as the {\it elementary} fragmentation function, since it represents an elementary fragmentation process from a single quark to a PS meson, and  in order to distinguish it from the {\it renormalized} one as discussed in Ref.~\cite{Ito:2009zc}. $\Delta$ corresponds to the relevant correlation for the fragmentation process. 
\section{Nonlocal chiral quark model (NLChQM) in Minkowski space}
In order to investigate this nonperturbative quantity, defined in Eq.~(\ref{eq:FRAG}), we will use NLChQM, which is motivated from the instanton-liquid model~\cite{Diakonov:1985eg,Shuryak:1981ff,Diakonov:1983hh,Diakonov:2002fq,Schafer:1996wv}. We note that, to date, various nonperturbative QCD properties have been well studied in terms of the instanton vacuum configuration, being comparable to experiments as well as lattice QCD simulations~\cite{Nam:2007gf,Nam:2010pt,Musakhanov:1998wp,Musakhanov:2002vu}. In the instanton model, nonperturbative QCD effects are deciphered by the nontrivial quark-instanton interactions in the dilute instanton ensemble, interpreting the spontaneous breakdown of chiral symmetry (SBCS). Considering these interesting features, below, we write the effective chiral action (EChA) of NLChQM in Minkowski space as follows:
\begin{equation}
\label{eq:ECA}
\mathcal{S}_\mathrm{eff}[m_f,\phi]=-\mathrm{Sp}\ln\left[i\rlap{/}{\partial}
-\hat{m}_f-\sqrt{M(\partial^2)}U^{\gamma_5}\sqrt{M(\partial^2)}\right],
\end{equation}
where $\mathrm{Sp}$ and $\hat{m}_f$ denote the functional trace $\mathrm{Tr}\int d^4x \langle x|\cdots|x\rangle$ over all the relevant spin spaces and SU(3) current-quark mass matrix $\mathrm{diag}(m_u,m_d,m_s)$, respectively. Note that, in deriving EChA in Eq.~(\ref{eq:ECA}), we have simply changed the Euclidean metric in the instanton-induced effective action into that for Minkowski space, assuming an appropriate analytic continuation between those spaces~\cite{Nam:2006au,Nam:2006sx}. Although this simple change of the metric is not fully understood, in practical applications, it works qualitatively very well~\cite{Praszalowicz:2001wy,Polyakov:1998td}. 

The momentum-dependent effective quark mass, which is generated from the nontrivial interactions between the quarks and nonperturbative QCD vacuum, can be written in a simple $n$-pole type form factor as follows:
\begin{equation}
\label{eq:MDM}
M(\partial^2)=M_0\left[\frac{n\Lambda^2}{n\Lambda^2-\partial^2+i\epsilon} \right]^{n}.
\end{equation}
Here, $n$ stands for an arbitrary but finite  positive integer. We will take $n=2$ with $\Lambda=600$ MeV, as usual in the instanton-motivated models~\cite{Nam:2006au,Nam:2006sx}, throughout the present work. Note that $U^{\gamma_5}$ stands for the nonlinear PS-meson field. For more details on this notation, one refers Refs.~\cite{Diakonov:1983hh,Diakonov:2002fq}. By expanding it up to $\mathcal{O(\phi)}$ from EChA in Eq.~(\ref{eq:ECA}), one can derive the following effective interaction Lagrangian density for the nonlocal quark-quark-PS meson vertex which is relevant for the fragmentation function in hand:
\begin{equation}
\label{eq:LAG}
\mathcal{L}_{\phi qq}=\frac{i}{F_\phi}\bar{q}\sqrt{M(\partial^2)}
\gamma_5(\lambda\cdot\phi)\sqrt{M(\partial^2)}q,
\end{equation}
where $F_\phi$ denotes the weak-decay constant for the PS meson $\phi$. 

Finally, after performing the trace over the relevant spin spaces in Eq.~(\ref{eq:FRAG}) using Eq.~(\ref{eq:LAG}), we arrive at a neat expression for the elementary fragmentation function in terms of the NLChQM as follows:
\begin{equation}
\label{eq:DDDDD}
d^\phi_{q,\bar{q}}(z,k^2_T,\mu)=\frac{1}{8\pi^3}\frac{M_kM_q}{2F^2_\phi}
\frac{z\{z^2k^2_T+[(z-1)\bar{M}_f+\bar{M}_{f'}]^2\}}
{[z^2k^2_T+z(z-1)\bar{M}^2_f+z\bar{M}^2_{f'}+(1-z)m^2_\phi]^2},
\end{equation}
where we have used the notation $\bar{M}_p=M_p+m_f=M(p^2)+m_f$, where $m_f$ indicates the current-quark mass with the flavor $f$. We note that, if we turn off the nonlocal interaction, i.e. taking $M_k=M_q=\mathrm{constant}$, the expression in Eq.~(\ref{eq:DDDDD}) turns into that for the local interaction models, such as Refs.~\cite{Bacchetta:2002tk,Amrath:2005gv,Ito:2009zc}, with a proper prefactor.  
\section{Numerical results}
\begin{figure}[h]
\begin{tabular}{cc}
\includegraphics[width=7.5cm]{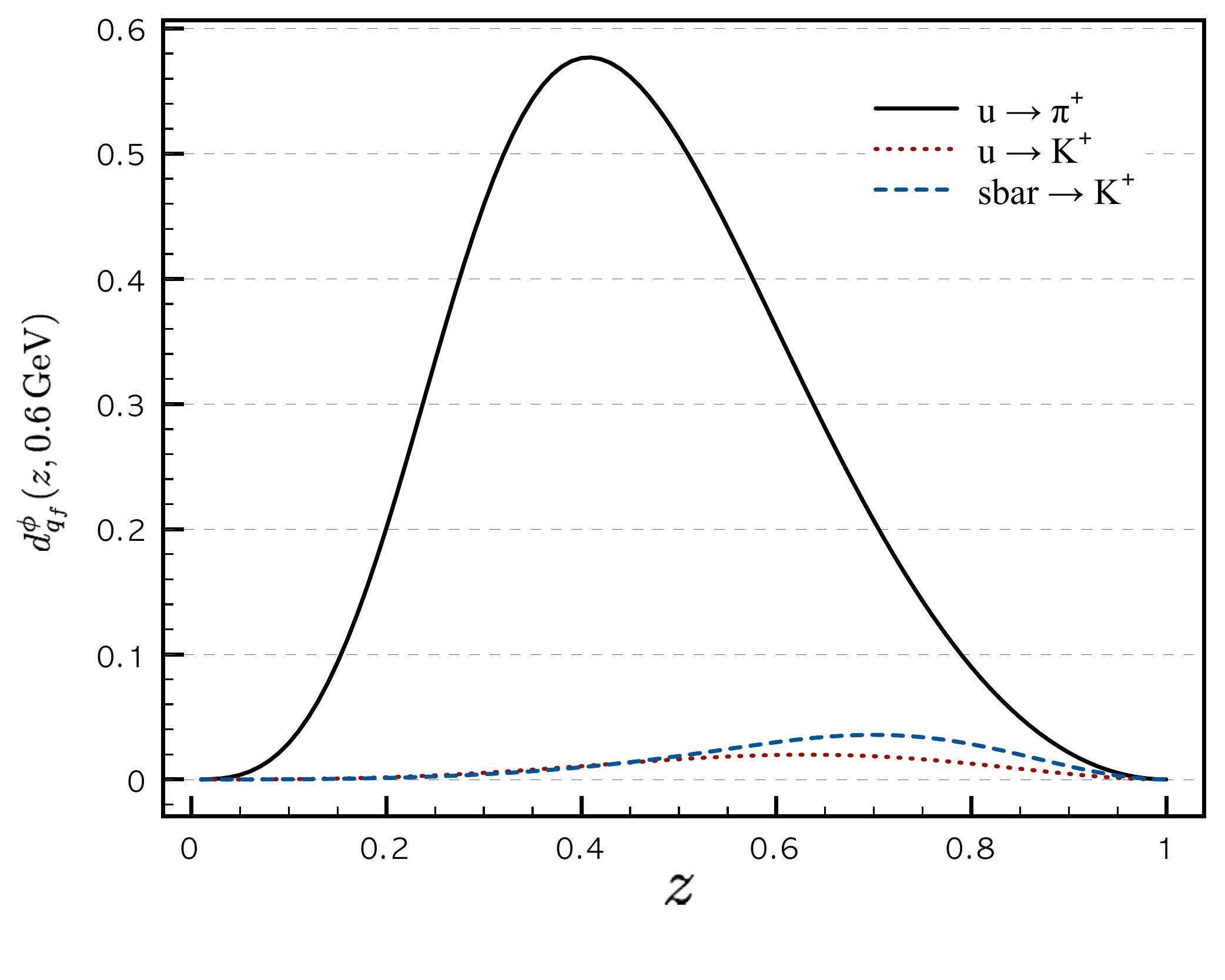}
\includegraphics[width=7.5cm]{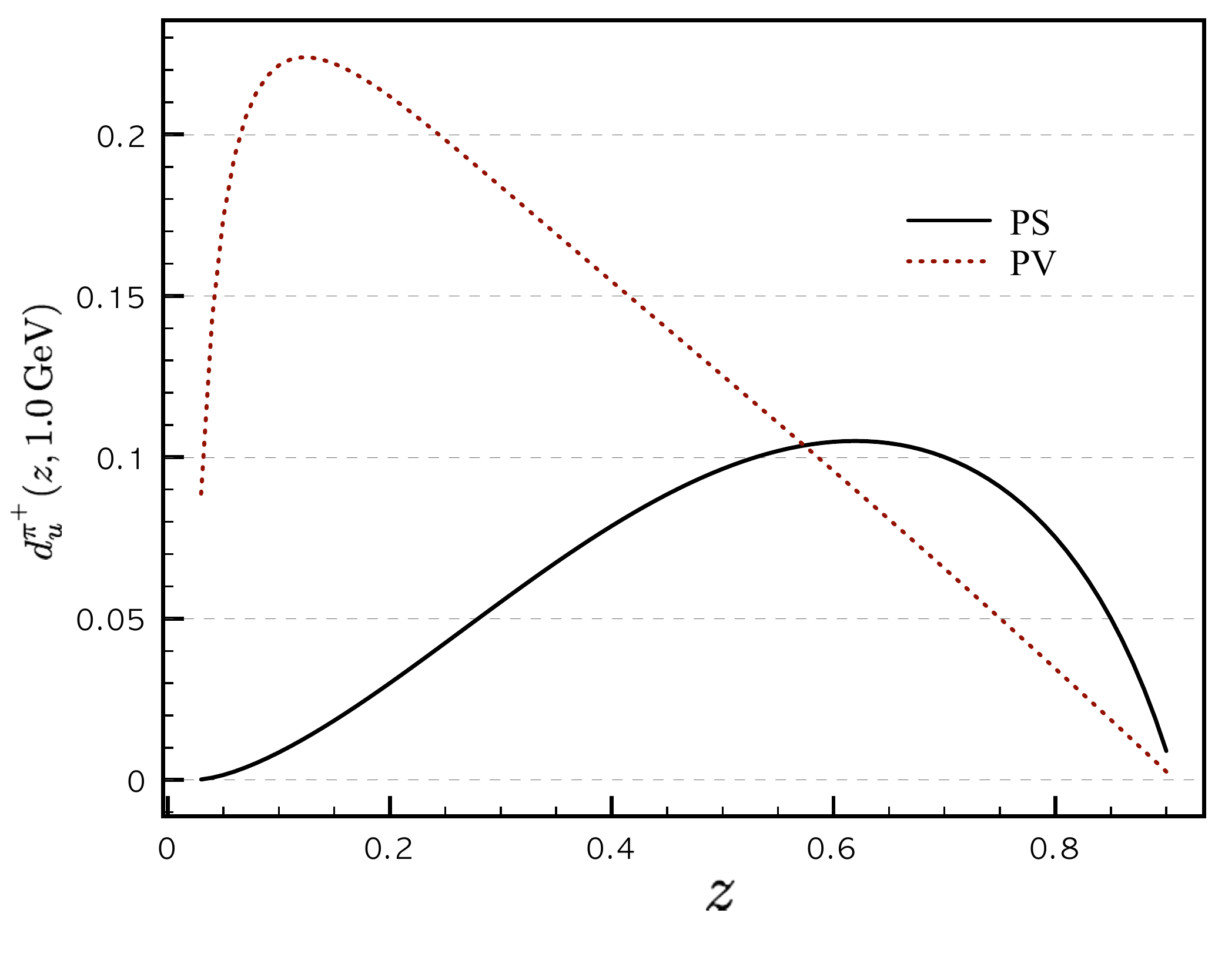}
\end{tabular}
\caption{(Color online) Favored elementary fragmentation functions for $u\to\pi^+$ (solid), $u\to K^+$ (dot), and $\bar{s}\to K^+$ (dash) at $\mu=0.6$ GeV (left). Those for $u\to\pi^+$ from the PS (solid) and PV (dot) schemes in Ref.~\cite{Bacchetta:2002tk} at $\mu=1.0$ GeV (right).}       
\label{FIG1}
\end{figure}

In this Section, we present numerical results for the elementary fragmentation function for the pion and kaon via the NLChQM. In the left panel of Fig.~\ref{FIG1}, we depict the curves for the favored elementary fragmentation functions as functions of $z$ at the renormalization scale $\Lambda\approx\mu=0.6$ GeV: $u\to\pi^+$ (solid), $u\to K^+$ (dot), and $\bar{s}\to K^+$ (dash).  As indicated in the left panel of Fig.~\ref{FIG1}, the kaon fragmentation functions are much smaller than that for the pion, since the pion fragmentation must be much populated at low energy scale $Q^2\approx\mu^2=(0.6\,\mathrm{GeV})^2$. In the right panel of Fig.~\ref{FIG1}, we show the numerical results from Refs.~\cite{Bacchetta:2002tk,Amrath:2005gv} for comparison. We note that the PS scheme result is almost the same with that from the NJL model calculation~\cite{Ito:2009zc}. If we compare ours with the PS-scheme result, which is comparable to the NJL one, it is found that the nonlocal quark-PS meson interaction makes considerable differences, i.e. the bump moves to smaller-$z$ regions, almost locating at the middle of the range $0\le z\le 1$.

Although, in this report, we do not perform the DGLAP evolution for higher $Q^2$ to compare with the empirical data for the fragmentation functions~\cite{Hirai:2007cx}, taking into account that the DGLAP evolution enhances the curve far more in the vicinity of $z=0$, the present numerical results with the bump in the middle are expected to give more reliable DGLAP evolution results, in comparison to those for the PS-scheme (or NJL model) results~\cite{Bacchetta:2002tk,Amrath:2005gv,Ito:2009zc}. Moreover, as expected, we can conclude that the present nonlocal results behave like the average between the PS- and PV-scheme ones qualitatively as shown in Fig.~\ref{FIG1}. More detailed analyses on these observations will be discussed in separate larger volumes.  

\section{Summary and Conclusion}
In this report, we have studied the unpolarized-fragmentation function, employing the nonlocal chiral quark model. From the numerical results, we observed that the nonlocal quark-PS meson interaction, which is a unique feature of the present theoretical framework, plays an important and distinctive role to produce the curves for the fragmentation function, in comparison to other effective model calculations. It was also viewed that calculated curves for the fragmentation function in the present work can provide more realistic results, being comparable to the empirical data after the DGLAP evolution to higher $Q^2$ values. More detailed analyses, including the meson-cloud effects, parton-distribution function using the Drell-Levi-Yan (DLY) relation, and DGLAP evolutions to higher $Q^2\,(\sim\mu^2)$, will appear soon elsewhere.
\begin{acknowledgements}
This talk was presented at the international conference {\it The Fifth Asia-Pacific Conference on  Few-Body  Systems in Physics 2011} (APFB2011), Seoul, Republic of Korea, $22-26$ August 2011. The authors thank P.~Ko, J.~W.~Chen, H.~Kohyama for fruitful discussions. S.i.N. is grateful to the hospitality during his staying at National Taiwan University (NTU) with the financial support from NCTS (North) in Taiwan, where the present work was initiated. The works of S.i.N. and B.G.Y. were partially supported by the grant NRF-2010-0013279 from National Research Foundation (NRF) of Korea. The work of C.W.K. was supported by the grant NSC 99-2112-M-033-004-MY3 from National Science Council (NSC) of Taiwan.
\end{acknowledgements}

\end{document}